\documentclass[reqno]{amsart}
\usepackage{amsmath,amsfonts,amsthm,amssymb,amsxtra}
\usepackage{graphicx}

\newtheorem{theorem}{Theorem}[section]
\newtheorem{proposition}[theorem]{Proposition}
\newtheorem{lemma}[theorem]{Lemma}
\newtheorem{corollary}[theorem]{Corollary}
\newtheorem{remark}[theorem]{Remark}

\theoremstyle{definition}

\numberwithin{equation}{section}

\newcommand{\be}{\begin{eqnarray}}
\newcommand{\ee}{\end{eqnarray}}
\newcommand{\bee}{\begin{eqnarray*}}
\newcommand{\eee}{\end{eqnarray*}}
\newcommand{\R}{{\Bbb R}}
\newcommand{\eps}{\varepsilon}
\newcommand{\C}{{\Bbb C}}
\newcommand{\N}{{\Bbb N}}

\begin{document}

\title[NLS with double-well potential]{Nonlinear Schr\"odinger equation with two symmetric point
interactions in one dimension}

\author {Hynek Kova\v{r}\'{\i}k}

\address {Dipartimento di Matematica, Politecnico di Torino}

\email {Hynek.Kovarik@polito.it}

\author {Andrea Sacchetti}

\address {Facolt\'a  di Scienze, Universit\'a di Modena e Reggio Emilia}

\email {Andrea.Sacchetti@unimore.it}


\date {\today}

\begin {abstract}
We consider a time-dependent one-dimensional nonlinear Schr\"odinger
equation with a symmetric potential double well represented by two
Dirac's $\delta$. Among our results we give an explicit formula for
the integral kernel of the unitary semigroup associated with the
linear part of the Hamiltonian. Then we establish the corresponding
Strichartz-type estimate and we prove local existence and uniqueness
of the solution to the original nonlinear problem.
\end{abstract}

\maketitle




\section {Introduction} \label{intro}

A presence of a certain symmetry in various physical systems is
often linked to some symmetric double well potential. Models with
such kind of potentials are therefore of interest in many fields of
research. One of the most important phenomena observed in these
situations is the spontaneous symmetry breaking, which arises in a
wide range of physical systems modelled by nonlinear equations with
double well potential. For instance, in classical physics it has
been experimentally observed for laser beams in Kerr media and
focusing nonlinearity \cite{Cambournac, Hayata}. Spontaneous
symmetry breaking phenomenon may also arise in the so-called Bose
Einstein condensates, where the effective double well potential is
formed by the combined effect of a parabolic-like trap and a
periodic optical lattice \cite {Albiez,Dalfovo,Raghavan}. From a
theoretical point of view, nonlinear systems with a double well
potential have been recently studied in the semiclassical limit
\cite {BambusiSacchetti,Sacchetti1} or in the limit of large
distance between the two wells \cite {Kirr,Sacchetti2}.

In order to introduce a simplified model of nonlinear Schr\"odinger
equations describing the basic features of systems with double-well
potentials, we consider in the present paper a one-dimensional
system where the two wells are represented by two Dirac's $\delta$
potentials.  In other words, we will deal with the nonlinear
time-dependent Schr\"odinger equation
\begin{equation} \label {nls}
\left \{
\begin {array}{l}
i\partial_t\psi_t = H_\alpha\psi_t + \nu\,  |\psi_t |^{2\mu }\, \psi_t  \\
\psi_t (x) |_{t=0} = \psi_0 (x),
\end {array}
\right.
\end{equation}
where $\psi_0$ is the initial data, $\mu,\nu$ are positive numbers
and $H_\alpha, \, \alpha\in\R$, denotes a one-dimensional
Schr\"odinger operator with symmetric delta interactions of strength
$\alpha$ and placed at the points $x=\pm a$, see section
\ref{prelim} below for a precise definition of $H_\alpha$. \ 
Although the stationary states and of the symmetry breaking
bifurcation for equation (\ref {nls}) with cubic nonlinearity have
been studied already by Jackson and Weinstein \cite {JW}, much less
is known about the time behaviour of the wave function for fixed
distance between the two wells.

The evolution equation \eqref{nls} with one delta interaction has been
considered by several authors
\cite{AdSa,DS,FOO,GHW,HMZ1,HMZ2,Sa,WMK}. In \cite{AdSa}, among other
things, local existence and uniqueness of the solution of the
associated nonlinear evolution equation is proved. The key point in
the analysis of the corresponding nonlinear problem in \cite{AdSa}
is a Strichartz type estimate for the time propagator of the free
Laplacian with one delta interaction restricted to the absolutely
continuous part of the spectrum. Such a Strichartz type inequality
follows, in the model with one delta interaction, easily from the
(quasi) explicit formula for the time propagator previously given by
\cite {ABD,ABD2,GS,S}.

Here we will deal with the situation when we have two delta
interactions of the same strength. As in \cite{AdSa} the crucial
ingredient is again a suitable Strichartz type estimate. The main
technical difficulty with respect to the case treated in \cite{AdSa}
is the fact that we don't have an explicit expression for the
associated time propagator at hand. Therefore, as a first step in
the analysis of equation \eqref{nls} we derive a formula for the
integral kernel of the time propagator of $H_\alpha$ in form of a
series of certain special functions, see Theorem \ref{kernel}. \ In
the next step we establish the corresponding Strichartz type
estimate, Theorem \ref{sqrt}. With the help of this Strichartz
estimate and standard tools of the analysis of nonlinear
Schr\"odinger equations, \cite{C}, we then prove local existence and
uniqueness of a solution of \eqref{nls}.

We would like to point out that our motivation was not only to
obtain the decay estimate \eqref{inf-1} for the evolution operator,
but also to give an explicit expression for the integral kernel of
the evolution operator, which can be used for numerical simulations
of equation \eqref{nls} based on spectral splitting methods, see
e.g. \cite {Sa} for numerical analysis of \eqref {nls} equation with
one single Dirac's $\delta$.

The text is organised as follows. After the preliminary section
\ref{prelim}, where we briefly recall some basic properties of the
operator $H_\alpha$, we announce the main results of our paper, see
section \ref{main-res}. Most of the proofs are presented in section
\ref{proofs}.  Some lengthy technical proofs are postponed to the
appendices. Concerning the notation,  we denote by $\|u\|_p$ the
norm of a function $u$ in $L^p(\R)$. In the case $p=2$ we will
sometimes write $\|u\|$ instead of $\|u\|_2$.


\section {Preliminaries}
\label{prelim}\noindent The operator $H_\alpha$ in $L^2(\R)$ is
associated to the closed quadratic form
\begin{equation}
Q[\psi] = \left \|\psi' \right \|^2 + \alpha \left ( | \psi (a)
|^2 + | \psi (-a)|^2 \right ), \quad \psi\in D(Q)=H^1(\R).
\end{equation}
It is well-known, see e.g. \cite{Al}, that the functions from the
domain of $H_\alpha$ satisfy the boundary conditions
\begin{align}
\psi (\pm a +0) &=\psi (\pm a -0)  \label {formula2}  \\
\alpha\, \psi (\pm a +0) &= \psi' (\pm a +0) - \psi' (\pm a -0) .
\label {formula3}
\end{align}
Then $H_\alpha$ acts on its domain $D(H_\alpha)$ as
$$
H_{\alpha} \psi = - \psi'' , \quad D(H_{\alpha} ) = \left \{ \psi
\in H^2 (\R \backslash \{ \pm a \} ) \, : \, \psi \, \, \, \mbox
{satisfies}\, \, \,  (\ref {formula2}) \ \mbox { and } \ (\ref
{formula3}) \right \} \, .
$$
Notice that, due to (\ref{formula2}), the function $\psi \in
D(H_{\alpha})$ is continuous at $x = \pm a $ and therefore
$D(H_{\alpha})$ is a subspace of $H^1 (\R)$. Thus, in the following
we denote by $\psi (\pm a )$ the limit (\ref {formula2}). It is
useful to recall some basic properties of the spectrum of
$H_{\alpha}$. The essential spectrum of $H_{\alpha}$ is purely
absolutely continuous and coincides with the positive real axis:
\bee
\sigma_{\mbox {\rm \small ess}} (H_{\alpha})=\sigma_{\mbox {\rm \small ac}}
(H_{\alpha}) =[0,+\infty ) \, .
\eee
From the explicit form of the resolvent operator
$(H_\alpha-z)^{-1}$, see \cite[Chap. II.2]{Al} it is straightforward
to determine the discrete spectrum of $H_\alpha$: if $\alpha  \geq
0$, then the discrete spectrum of $H_{\alpha}$ is empty. For $\alpha
< 0$ the discrete spectrum consists of negative eigenvalues $E$
given given by the implicit equation
\bee
(-2 i k +\alpha )^2 = \alpha^2 e^{i 4 k a }, \ \ k = \sqrt {E} , \quad  \Im k \geq 0 \, .
\eee
In particular, if $ a  \le - \frac {1}{\alpha}$ then the discrete
spectrum of $H_{\alpha}$ consists of only one eigenvalue $E_1 (a
,\alpha)$ defined by
\begin{equation}
E_1 (a,\alpha) = - \frac {1}{4a^2} \left [ W \left ( - a \alpha e^{a
\alpha } \right )
 - a \alpha \right ]^2 \, , \label {Thm2a}
\end{equation}
where $W(\cdot)$ is the Lambert's special function defined by the
equation $W(x) e^{W(x)} = x $, \cite {CGHJK}. If $ a  > -\frac
{1}{\alpha}$ then the discrete spectrum of $H_{\alpha}$ consists of
two eigenvalues: $E_1 (a , \alpha )$ and
\begin{equation}
E_2 (a,\alpha ) = - \frac {1}{4a^2} \left [ W \left ( + a \alpha e^{a \alpha } \right ) - a
\alpha \right ]^2 \, .\label {Thm2b}
\end{equation}


\section{Main results}
\label{main-res}

We first state a result which gives an explicit formula for the evolution
operator associated with $H_\alpha$ in terms of a series of
parabolic cylinder functions. We then make use of this formula to
study local existence and uniqueness of the nonlinear problem
\eqref{nls}. In order to formulate our results we need to introduce
some notation. Let $P_c$ be the spectral projector of $H_\alpha$ on
its absolute continuous spectrum. Moreover, define
\begin{align*}
z_n := z_n (x,y)  = 4 a n + & |x+a|  + |y+a|\\
 a_n  := a_n (x,y,t) =
\frac{z_n}{4\sqrt{t}}- \frac{\alpha\sqrt{t}}{4}\, , \qquad & b_n :=
b_n (x,y,t) = \frac{z_n}{4\sqrt{t}}+ \frac{\alpha\sqrt{t}}{4} \, .
\end{align*}

\begin{theorem} \label{kernel}
Let $t>0$ and assume that $a\alpha \not= -1$.  Then the integral
kernel of the operator $\exp (-it\, H_\alpha)\, P_c$ is given by
\be \label{kernel:eq} U_\alpha(t;x,y) =  \frac {1}{\sqrt {4\pi i t}
}\, \, e^{i|x-y|^2/4t} - \frac {1}{2\pi} \sum_{m=0}^{+\infty} \left
[ p_m(t;x,y) + p_m(t;-x,-y) \right ], \ee
where
\bee
p_m(t;x,y) = (-1)^m\, r_{\frac {m}{2}} \left (t; x , (-1)^m y \right )\, ,
\eee
\begin{align}
r_n(t;x,y) & =  \mbox {\rm sign} (\alpha ) \sqrt {\frac {\pi}{2}}\,
|\alpha|^{2n+1} \left ( \frac {it}{2} \right )^n e^{i \frac
{z_n^2}{4t}}  e^{(ia_n - b_n )^2} D \left [ -2n -1 ,-\mbox {\rm
sign} (\alpha )2i (a_n+ ib_n) \right ] \label {eq.pippo}
\end{align}
and $D[n,z]$ denotes the parabolic cylinder function.  Moreover, the
series on the right hand side of \eqref{kernel:eq} converges for any
$t>0$ uniformly with respect to $x$ and $y$.
\end{theorem}

\begin {remark}
Here we assume $a \alpha \not= -1$ for technical reasons. Note that
for $a \alpha =-1$ the second eigenvalue $E_2$ of $H_\alpha$ emerges
from the continuous spectrum.
\end {remark}

\begin{remark} The evolution operator $e^{-i t H_\alpha}$ then has the kernel
\bee {\mathcal V}_\alpha(t;x,y) = \Theta (-\alpha ) e^{-i t E_1} \,
\bar \varphi_1\,  (y) \varphi_1 (x) + \Theta (-1-a \alpha ) e^{-i t
E_2} \, \bar \varphi_2 (y)\, \varphi_2 (x) + U_\alpha(t;x,y) \eee
where $\varphi_1$ and $\varphi_2$ are the normalised eigenfunctions
associated to the eigenvalues $E_1$ and $E_2$ and $\Theta(\cdot)$
denotes the Heaviside function.
\end {remark}

\begin{remark}
Time evolution in the models involving one or more delta
interactions similar to ours have been studied, apart from
\cite{AdSa}, in many other works, see e.g. \cite{clr00,clr02,cls}.
For example in \cite{clr00} is considered a model of ionization in
which to the unperturbed Hamiltonian $-\frac{d^2}{dx^2}
+g\delta(x)$, with $g>0$ is applied a perturbing time dependent
potential of the form $-g \eta(t) \delta(x)$, where $\eta(t)$ is a
periodic function of time. It  is then shown that the survival
probability of the bound states of the unperturbed Hamiltonian tends
to zero as $t\to\infty$ and the nature of this decay is discussed in
detail. Similar survival probability is studied in \cite{clr02} for
a model where the perturbing potential is of the form
$\eta(t)(\delta(x+a)-\delta(x-a))$ with $\eta(t)=r \sin (\omega t)$.

In our model the linear part of the of the Schr\"odinger equation
corresponds to the Hamiltonian $H_\alpha$ with time-independent
potential represented by two symmetric delta interactions.
Therefore, instead of looking at the time evolution of the
(possible) bound states, we focus on the time evolution of the
"continuous part" of the operator, i.e. on the evolution operator
restricted to the absolutely continuous part of the spectrum.
\end{remark}

\noindent  As a consequence of Theorem \ref{kernel} we obtain the
following dispersion estimate

\begin{corollary} \label{sqrt}
Assume that $a\alpha \not= -1$. Then for any there exists a constant $C$ such
that for any $u\in L^1(\R)$ and any $t>0$ it holds
\begin{equation} \label{inf-1}
\| e^{-i t H_\alpha}P_c\, u\|_\infty \, \leq \, C\, t^{-\frac 12}\,
\|u\|_1.
\end{equation}
\end{corollary}

\begin{remark} \label{regular}
The same decay behavior for the model with one $\delta$ interaction
was obtained in \cite{AdSa}. Dispersion estimates for Schr\"odinger
operators with regular potentials have been well studied in the
literature, see for example \cite{ns,wed}. In \cite{wed} it was
shown, under certain regularity assumptions on the corresponding
potential, that the decay rate $t^{-1/2}$ of the unitary group
$e^{-itH}$ as an operator from $L^1(\R)$ to $L^\infty(\R)$ is
typical for one-dimensional Schr\"odinger operators.

Although the assumptions of \cite{ns,wed} do not allow to include
directly the situation treated in our model, it is reasonable to
expect that the dispersion estimate \eqref{inf-1} could also be
obtained by approximating the delta interactions by suitable
sequences of regular potentials and applying the results of
\cite{wed}.
\end{remark}

\noindent Our main result concerning the evolution problem \eqref{nls} reads as follows

\begin{theorem} \label{nls:thm}
Suppose that the assumptions of Theorem \ref{kernel} are satisfied.
If $\psi_0\in H^1(\R)$, then the problem \eqref{nls} admits a unique
local solution
\begin{equation} \label{solution}
\psi_t \in C((0,T),\, H^1(\R))\, \cap\, C^1((0,T), \, H^{-1}(\R))
\end{equation}
for some $T>0$. Moreover, if $\psi_t$ is a solution of \eqref{nls},
then
\begin{equation} \label{energy-consev}
\|\psi_t\| = \|\psi_0\|, \qquad
{\mathcal E} [\psi_t ] = {\mathcal E} [\psi_0] \qquad t\in (0,T),
\end{equation}
where ${\mathcal E} $ is the energy functional on $H^1(\R)$ given by
\begin{equation}
{\mathcal E} [\psi ] = \left \|\partial_x\psi \right \|^2 + \alpha
\left [ | \psi (a) |^2 + | \psi (-a)|^2 \right ] + \frac {\nu}{\mu
+1}\,  \| \psi \|^{2\mu +2}_{2\mu +2}\, .\label {En}
\end{equation}
\end{theorem}

\section{Proofs of the main results}
\label{proofs}
We will need the explicit form of the integral kernel of the resolvent of $H_\alpha$.
According to \cite[Chap.II.2]{Al}, we have
\bee
\left ( \left [ H_{\alpha } - k^2 \right ]^{-1} \phi \right )(x) =
\int_{\R} K_\alpha (x,y;k) \phi (y) d y,
\quad \phi\in L^2(\R) , \ \Im k \ge 0 \, ,
\eee
where the integral kernel $K_\alpha$ is given by
\be \label{resolvent}
K_\alpha (x,y;k) = K_0 (x,y;k) + \sum_{j=1}^4 K_\alpha^j (x,y;k)
\ee
with
\be K_\alpha^j (x,y;k) & = - \left(2k((2k+i \alpha )^2 + \alpha^2
e^{i4ka})\right)^{-1}L^j_\alpha(x,y;k)\, , \, \, \, \, K_0 (x,y;k)
= \frac {i}{2k}\,  e^{ik|x-y|} \label{resolvent-K} \ee
and
\begin{align}
L_\alpha^1 (x,y;k) & = - \alpha(2k+i\alpha)\,  e^{ik |x+a|} e^{ik|y+a|}, &
L^4_\alpha(x,y;k)  = L_\alpha^1 (-x,-y;k) \nonumber \\
L_\alpha^2 (x,y;k) & = i\alpha^2\, e^{2ika}\,  \, e^{ik |x+a|} e^{ik|y-a|}, &
L^3_\alpha(x,y;k)  = L_\alpha^2 (-x,-y;k) . \label{resolvent-bis}
\end{align}
%


\subsection {Evolution operator of the linear equation}
\label{evol} \noindent We split the proof of Theorem \ref{kernel}
into several lemmas. We start by studying the properties of the
function $U_\alpha(t;x,y)$ defined by
\be
\label{evol:integral}
U_\alpha(t;x,y)  = - \frac {i}{\pi } \int_{\R} k\, e^{-i k^2 t} K_\alpha(x,y;k) dk \, .
\ee
\noindent Note that
\be
U_\alpha (t;x,y) = U_0 (t;x,y) + \frac {i}{2\pi} \int_\R  e^{-ik^2t}
\frac {f_\alpha (x,y;k)}{(2k+i\alpha)^2 + \alpha^2 e^{i4ka}} dk \label {eq311}
\ee
with
\bee f_\alpha (x,y;k) = \sum_{j=1}^4 L_\alpha^j (x,y;k), \quad
U_0(t;x,y) =  \frac {1}{\sqrt {4\pi i t}}\, \, e^{i|x-y|^2/4t}.
\eee

\begin{lemma} \label{interchange}
For any $t>0$ it holds
\be U_\alpha (t;x,y) = U_0(t;x,y) - \frac {1}{2\pi }\, \sum_{j=1}^4
U_j(t;x,y) \label{evol-u}, \ee
where
\begin{align}
U_j(t;x,y) &=  -i\, \sum_{n=0}^\infty\, \int_{\R} e^{-i k^2 t}
\frac{L^j_\alpha(x,y;k)}{(2k+i\alpha)^2}\, \left(-\frac{\alpha^2\,
e^{i4ka}}{(2k+i\alpha)^2} \right)^n\, dk  \label{u1}.
\end{align}
\end{lemma}
\begin{proof}
By \eqref {eq311} we have
\begin{align}
U_\alpha(t;x,y) - U_0(t;x,y) & =  \frac{i}{2\pi} \int_{\R } e^{-i
k^2 t}\, \frac{f_\alpha(x,y;k)}{(2k+i \alpha )^2 + \alpha^2
e^{i4ka}}\, dk
\nonumber \\
& =  \frac{i}{2\pi} \int_{\R} e^{-i k^2 t} \,  \sum_{n=0}^\infty\,
\frac{f_\alpha(x,y;k)}{(2k+i\alpha)^2}\,
 \left(-\frac{\alpha^2\, e^{i4ka}}{(2k+i\alpha)^2} \right)^n\, dk . \label{series}
\end{align}
Since
\bee f_\alpha(x,y;k) = \mathcal{O}(k) \qquad k \to 0, \eee
for each $x,y$, the series on the right hand side of \eqref{series}
converges uniformly with respect to $k$ on $\R$. We can thus
interchange the summation and the integration to get \eqref{u1}.
\end{proof}

\begin {lemma} \label{AlphaPos1}
For $t>0$ and  $a\alpha \not= -1$ we have
\bee U_1(t;x,y) = \sum_{n=0}^\infty r_n(t;x,y) , \quad U_2(t;x,y) =
\sum_{n=0}^\infty \hat r_{n+\frac 12}(t;x,y), \quad \hat r_n(t;x,y)
= - r_n(t;x,-y) . \eee
\end {lemma}

\begin {proof}
By Lemma \ref{interchange}
\be
U_1(t;x,y) &=& \, \sum_{n=0}^\infty i^{2n+1} \alpha^{2n+1} \int_{\R} e^{-i k^2 t}
\frac {e^{i4nka}}{(2k+i\alpha)^{2n+1}}
e^{ik |x+a|} e^{ik|y+a|} dk \label {formula5}  \\
&=&  \sum_{n=0}^\infty i^{2n+1} \alpha^{2n+1} e^{\frac{iz_n^2}{4t}}
2^{-(2n+1)} t^n \int_{\R } e^{-i r^2 } \left
(r+\frac{z_n}{2\sqrt{t}}+\frac{i\alpha \sqrt {t}}{2} \right )^{-(2n+1)} dr \nonumber \\
&=&  \sum_{n=0}^\infty i^{2n+1} \alpha^{2n+1} e^{\frac{iz_n^2}{4t}} 2^{-(2n+1)} t^n A_{2n+1}
\left ( \frac{z_n}{2\sqrt{t}},\,  \frac{\alpha \sqrt {t}}{2} \right ) \nonumber  \\
&=& \sum_{n=0}^\infty \mbox {sign} (\alpha ) \sqrt {\frac {\pi}{2}}\, \, \alpha^{2n+1}
\left ( \frac {it}{2} \right )^n e^{\frac {i z_n^2}{4t}}\,  e^{-
(a_n + i b_n )^2} D \left [ -2n -1 ,-\mbox {sign} (\alpha ) 2i (a_n + i b_n ) \right ] \nonumber
\ee
where we apply formula (\ref {formula11}), see appendix
\ref{technical}. The calculation of $U_2(t;x,y)$ follows the same
line. Let $ \hat z_n :=\hat z_n (x,y) = z_n (x,-y)$ and define
\bee
\hat a_n  :=\hat a_n (x,y,t)= \frac{\hat z_n}{4\sqrt{t}}-
\frac{\alpha\sqrt{t}}{4} \ \ \mbox { and } \ \ \hat b_n :=\hat b_n
(x,y,t) = \frac{\hat z_n}{4\sqrt{t}}+ \frac{\alpha\sqrt{t}}{4} \, .
\eee
As above we obtain
\bee
U_2(t;x,y) &=&  - \sum_{n=0}^\infty\, i^{2n+2} \alpha^{2n+2}
\exp\left(i\left( \hat z_{n+\frac 12}\right )^2/4t \right)\,
2^{-(2n+2)}\, t^{n+\frac 12}
A_{2n+2} \left ( \frac{\hat z_{n+\frac 12}}{2\sqrt {t}},\,  \frac{\alpha \sqrt {t}}{2}  \right ) \\
&=& - \sum_{n=0}^\infty \mbox {sign} (\alpha ) \sqrt {\frac
{\pi}{2}}\, \, \alpha^{2n+2} \left ( \frac {it}{2} \right )^{\frac
{2n+1}2} \exp\left(i\left( \hat z_{n+\frac 12}\right )^2/4t \right)\,
e^{- \left(\hat a_{n+\frac 12} + i \hat b_{n+ \frac 12} \right)^2} \\
& & \qquad \quad D \left [ -2n -2 ,- \mbox {sign} (\alpha )2i (\hat
a_{n+\frac 12} + i \hat b_{n+\frac 12} ) \right ]
\eee
where we again applied formula (\ref {formula11}).
\end {proof}

\noindent We will also need a uniform estimate on the sequence $r_n(t;x,y)$ defined in \eqref{eq.pippo}.

\begin {lemma} \label{rest-term}
Let $t>0$ be fixed and assume that $a\alpha \not= -1$. Then there
exists a constant $C$ independent of $t$, $x$ and $y$, such that for
all $n\geq 1$ we have
\be
|r_n(t;x,y) | ,\, |r_{n+\frac 12}(t;x,y) | \, \leq \, C^n t^{\frac 12}\, \,
\frac{t^{2n}\, n^{-2n}}{(1+|x+a|+|y+a|)^{2n}}
\label {stimarn}
\ee
\end {lemma}

\begin {proof}
From the definition of $a_n, b_n$, Lemma \ref{LemmaA}, appendix
\ref{technical}, and Lemma \ref{AlphaPos1} we get
\be
\label{rn}
|r_n(t;x,y )| \, \leq\,  \sqrt {\frac {\pi}{2}}\, \, |\alpha|^{2n+1}
\left ( \frac {t}{2} \right )^n\,  |I_n(t;x,y)| ,
\ee
where
\bee
I_n (t;x,y) = \int_{\R} e^{- s^2} (s+ \sqrt{2}\, a_n+ i\sqrt{2}\, b_n )^{-2n-1} ds
\eee
Hence
\bee
|I_n (t;x,y)| \, \leq \, \sqrt{\pi}\, |\sqrt{2}\, b_n|^{-2n-1} \leq \,
C\, \left(\frac{t}{z_n}\right)^n,
\eee
which implies \eqref{stimarn}. The estimate for $|r_{n+\frac
12}(t;x,y) |$ is completely analogous.
\end {proof}


\noindent Next we show that $U_\alpha(t;x,y)$ defines the kernel of the
evolution operator associated to $H_\alpha$.

\begin {lemma} \label{evol-lemma}
Let $K_\alpha (x,y;k)$ be the kernel of the resolvent operator of
$H_\alpha$ and let  $P_c$ be the spectral projection of $H_\alpha$ on
$[0,\infty)$. Then for any test
function $\phi\in C_0^\infty (\R )$ we have
\be \label{evol:eq}
\left ( e^{-itH_{\alpha} } P_c\,  \phi \right ) (x) =
\int_{\R} U_\alpha(t;x,y) \, \phi (y)\, d y.
\ee
\end {lemma}

\begin {proof} Fix $\phi\in C_0^\infty (\R )$. By \cite[Thm.3.1]{T} it holds
\begin{equation} \label{s-limit}
\lim_{\epsilon \to 0^+} e^{-(i t+\epsilon) H_\alpha } \phi =
e^{-i t H_\alpha } \phi \, ,
\end{equation}
Since the absolute continuous spectrum of $H_\alpha$ is the interval
$[0,\infty )$, the spectral  theorem gives
\bee
e^{-(i t+\epsilon) H_\alpha } P_c = \int_0^\infty e^{-(i t+\epsilon) \lambda } d P_\lambda ,
\eee
where $P_\lambda$ is the spectral projector of $H_\alpha$ on the interval $[0,\lambda]$.
By the Stone's formula, see, e.g. \cite[ThmVII.13]{RS}, we have that for any $\lambda \ge 0$
\bee
P([0,\lambda]  ) &=& \lim_{\delta \to 0^+} \frac {1}{2\pi i} \int_0^\lambda
\left \{ [H_\alpha -(z+i\delta)  ]^{-1} -  [H_\alpha -(z-i\delta) ]^{-1} \right \} dz\, .
\eee
Hence
\begin{align}
 e^{-(i t+\epsilon) H_\alpha } P_c\, \phi  &= \frac
{1}{2\pi i} \int_0^\infty e^{-(i t+\epsilon) z } \lim_{\delta \to
0^+} \int_\R \left \{ K_\alpha \left ( x,y;\sqrt {z+i\delta } \right
) - K_\alpha
\left ( x,y;\sqrt {z-i\delta } \right ) \right \} \phi (y) d y d z \nonumber \\
& \ = \frac {1}{\pi i} \int_{-\infty}^0 e^{-(i t+\epsilon) k^2 }
\lim_{\delta \to 0^+}  \int_\R k\, K_\alpha \left ( x,y; \sqrt
{k^2-i\delta }\right ) \phi (y)\, d y\, d k  \nonumber \\
& \qquad + \frac {1}{\pi i} \int_0^{\infty} e^{-(i t+\epsilon) k^2 }
\lim_{\delta \to 0^+}  \int_\R k\, K_\alpha \left ( x,y; \sqrt
{k^2+i\delta }\right ) \phi (y)\, d y\,  d k , \label{interm}
\end{align}
where we have substituted $k=\sqrt {z}$ in the integral containing
the kernel $K_\alpha \left ( x,y;\sqrt {z+i\delta } \right )$ and
$k=-\sqrt {z}$ in the integral containing the kernel $K_\alpha \left
( x,y;\sqrt {z-i\delta } \right )$. we find out that. Since $\phi\in
C_0^\infty (\R )$ and $|k K_\alpha(x,y;\sqrt {k^2+i\delta })|$ is
uniformly bounded for $\delta\geq 0$ small enough, we can exchange
the limit and integration in \eqref{interm}. Note also that
\bee
\lim_{\delta \to 0^+}  k K_\alpha \left ( x,y; \sqrt {k^2\pm i\delta
} \right ) = k\,  K_\alpha (x,y; \pm |k|)\qquad  \forall\, k\neq 0
\eee
and that $k K_\alpha (x,y;k )= \mathcal{O}(1)$ as $k\to 0$. From the
Fubini theorem we then obtain
\begin{equation}
\left ( e^{-(it+\epsilon) H_\alpha} P_c \phi \right ) (x) = \frac
{1}{\pi i} \int_{\R }  \phi (y) \int_{\R} k K_\alpha (x,y,k) e^{-
k^2(it+\epsilon)} \, d k \, dy  = \int_{\R} U_\alpha
(t-i\epsilon;x,y) \phi (y) d y .\label{evol-eps}
\end{equation}
Passing to the limit $\epsilon \to 0+$ in \eqref{evol-eps} gives
\begin{equation} \label{lim-int}
\left ( e^{-itH_\alpha} P_c\, \phi \right ) (x) = \lim_{\epsilon\to
0+}\, \int_{\R} U_\alpha (t-i\epsilon;x,y) \phi (y) d y \,  ,
\end{equation}
see \eqref{s-limit}. By  \eqref{eq311} it can be directly verified
that for any $x,\, y\in\R$ it holds
\begin{equation}
\lim_{\epsilon\to 0+}\, U_\alpha (t-i\epsilon;x,y) = U_\alpha
(t;x,y) \label {unif}.
\end{equation}
Indeed, since $U_0(t;x,y)$ is continuous in $t\in \C\setminus\{0\}$,
it follows that
\begin{align} \label{contin}
U_\alpha (t;x,y) - U_\alpha (t-i\epsilon;x,y) & = -\frac{i}{\pi}
\int_{\R } e^{-i k^2 t}\, \frac{f_\alpha(x,y;k)(1-e^{-\epsilon
k^2})}{(2k+i \alpha )^2 + \alpha^2 e^{i4ka}}\, \, dk + o(1)
\end{align}
as $\epsilon\to 0+$. To show that the first term on the right hand
side of \eqref{contin} converges to zero, we observe that
\begin{equation} \label{R-finite}
\lim_{\epsilon\to 0+}\, \int_{-R}^R e^{-i k^2 t}\,
\frac{f_\alpha(x,y;k)(1-e^{-\epsilon k^2})}{(2k+i \alpha )^2 +
\alpha^2 e^{i4ka}}\, \, dk =0
\end{equation}
for any $R>0$. On the other hand, a direct calculation shows that
\begin{align*}
\frac{f_\alpha(x,y;k)}{(2k+i \alpha )^2 + \alpha^2 e^{i4ka}} &=
-\frac{\alpha}{2k}\, \left( e^{ik|x+a|}\,
e^{ik|y+a|}+e^{ik|x-a|}\,e^{ik|y-a|}\right) +\mathcal{O}(k^{-2}),
\end{align*}
where the term $\mathcal{O}(k^{-2})$ depends also on $x,y$ and $\alpha$. From the asymptotic relation
$$
 \int_{R}^\infty e^{-z
k^2 }\, \frac 1k \, dk = \mathcal{O}\left(\frac{e^{-z R^2}}{zR^2}\right), \quad R\to\infty,
\quad |\mbox{arg\, }z| < \pi,
$$
see e.g. \cite[Chap.5]{AbSt}, we thus deduce that
\bee
\int_{R}^\infty e^{-z k^2 }\, \frac{f_\alpha(x,y;k)(1-e^{-\epsilon k^2})}{(2k+i \alpha
)^2 + \alpha^2 e^{i4ka}} = \mathcal{O}(R^{-2}), \quad R\to \infty
\eee
uniformly with respect to $\epsilon\geq 0$. The integral
corresponding to $k\in (-\infty,-R)$ is treated in the same way.
This together with \eqref{R-finite} implies \eqref{unif}. It follows
that for any $x\in\R$ the function $|U_\alpha (t-i\epsilon;x,y)|$ is
uniformly bounded for $\epsilon$ nonnegative and small enough and
$y$ on compact subsets of $\R$. Hence we can interchange the limit
with the integral in \eqref{lim-int} to conclude the proof.
\end {proof}

\begin{proof}[Proof of Theorem \ref{kernel}]
Since
\be
U_3(t;x,y) =  U_2(t;-x, -y), \quad U_4(t;x,y) =  U_1(t;-x, -y)\, , \label{u34}
\ee
the statement of the Theorem follows from Lemmata \ref{interchange},
\ref{AlphaPos1}, \ref{rest-term} and \ref{evol-lemma}.
\end {proof}


\subsection{Proof of Corollary \ref{sqrt}}

\noindent We need two auxiliary Lemmata.
\begin{lemma}  \label{kernel-estim}
Let $a\alpha\neq -1$. For any $T>0$ there exists a constant $C_T$ such that
\bee
\sup_{x,y\in\R}\, \left | U_\alpha (t;x,y)\right | \leq \, C_T\, \, t^{-\frac 12},\qquad t\in[0, T].
\eee
\end{lemma}

\begin{proof}
Clearly
\bee \left | U_0 (t;x,y)\right | \leq \frac{1}{\sqrt{4\pi t}}. \eee
It thus remains to estimate $U_j(t;x,y), j=1,2,3,4$.  By
\eqref{stimarn} we have
\bee
\left | \sum_{n=1}^\infty r_{n+\frac 12}(t;x,y) \right|\, , \, \left
| \sum_{n=1}^\infty r_n(t;x,y) \right| \,
\leq \, C'\, t^{\frac 12}, \qquad t\in[0,T],
\eee
where the constant $C'$ depends only on $T$.  On the other hand, for
$n=0$ it follows directly from the definition that
\bee
| r_0(t;x,y)| \, ,\, |r_{\frac 12}(t;x,y)| \leq C,
\eee
where $C$ is a real number.
\end{proof}

\noindent Analogous upper bound on $U_\alpha(x,y;t)$ for large $t$
is given in the following Lemma, whose proof we postpone to appendix
\ref{long-time}.

\begin{lemma} \label{large-time}
There exists a constant $C$ such that for all $t$ large enough it holds
\be \label{large:eq}
\sup_{x,y\in\R }\, |\sqrt{t}\, \, U_\alpha(x,y;t)| \, \leq \, C.
\ee
\end{lemma}

\begin{proof}[Proof of Corollary \ref{sqrt}]
By Lemmata \ref{kernel-estim} and \ref{large-time} we have
\bee \sup_{x,y\in\R}\, \left | U_\alpha (t;x,y)\right | \leq \, C \,
t^{-\frac 12}
\eee
for all $t>0$ and some constant $C$ independent of $t$. Hence
\bee
\| e^{-i t H_\alpha}P_c\, \phi\|_\infty \, \leq \, C\, t^{-\frac 12}\,
\|\phi\|_1 \qquad \forall\, \phi\in C_0^\infty(\R).
\eee
By density, this inequality extends to all $u\in L^1(\R)$.
\end{proof}

\noindent  As a consequence we obtain the following Strichartz-type estimate.

\begin{corollary} \label{str}
Let $r\geq 2$ and $q=\frac{4r}{r-2}$. Then for any $T>0$ there
exists a constant $C$ such that
\begin{equation} \label{str:eq}
\| e^{-i t H_\alpha}P_c \, u\|_{L^q((0,T),L^r(\R))}\, \leq \, C\,
\|u\|_2 \qquad \forall\, u\in L^2(\R).
\end{equation}
\end{corollary}

\begin{proof}
Let $0< t< T$. In view of \eqref{inf-1} and the obvious inequality
\bee
\|e^{-i t H_\alpha}P_c\,  u\|_2 \, \leq \, \|u\|_2,
\eee
the Riesz-Thorin interpolation theorem implies that
\begin{equation} \label{interp}
\|e^{-i t H_\alpha}\, P_c\, u\|_p \, \leq \, C \, t^{\frac 1p-\frac 12}\,
\|u\|_r, \quad \frac 1p+\frac 1r =1
\end{equation}
for some $C$ and $p\in[2,\infty]$. Since $H_\alpha$ is self-adjoint,
\eqref{str:eq} follows from \eqref{interp} and \cite[Thm. 2.7.1]{C}.
\end{proof}

\vspace{0.1cm}


\subsection{Proof of Theorem \ref{nls:thm}}

\noindent In order to study the well-posedness of the nonlinear
equation \eqref{nls}, we follow the strategy adopted in \cite{AdSa}.
To this end we define the operator $T_\alpha$ by
\bee
T_\alpha =  \left \{
\begin {array}{ll}
H_\alpha &    \quad \alpha \geq 0  \\
 H_\alpha -E_1(a,\alpha)  &  \quad \alpha <0.
\end {array}
\right.
\eee
The key point, apart from the Strichartz estimate, is to prove the following technical
result, see also \cite[Sec.3.7]{C}.

\begin{proposition} \label{aux}
Let $p\in[2,\infty]$. Then the domain $D(T_\alpha)$
of $T_\alpha$ is embedded in $L^p(\R)$. Moreover, there exists a
constant $C$ such that for any $\eps<0$ and any $u\in L^p(\R)$ it
holds
\begin{equation} \label{lp:estim}
\|(\eps\, T_\alpha-1)^{-1}\, u\|_{p}\, \leq \, C\, \|u\|_{p} .
\end{equation}
\end{proposition}

\begin{proof}
Consider first the case $\alpha>0$. \ Since $H^1(\R)$ is
continuously embedded in $L^p(\R)$ for any $p\in[2,\infty]$, the
embedding $D(T_\alpha)=D(H_\alpha) \hookrightarrow L^p(\R)$ follows
in view of the fact that $D(H_\alpha)$ is a subspace of $H^1(\R)$.
To prove \eqref{lp:estim} we note that $(\eps\,
T_\alpha-1)^{-1}=(\eps\, H_\alpha-1)^{-1}$ is an integral operator
with the kernel
\bee
\eps^{-1} K_\alpha(x,y;i\lambda) =  \eps^{-1} K_0(x,y;i\lambda) +
\eps^{-1} \sum_{j=1}^4 K^j_\alpha(x,y;i\lambda), \quad \lambda= -i\,
\eps^{-\frac 12} >0.
\eee
First we observe that
\bee
\eps^{-1} \int_\R K_0(x,y;i\lambda) u(y)\, dy = \frac{1}{2\lambda\eps}\,
 \left( e^{-\lambda |\cdot|} \ast u \right) (x).
\eee
By the Young inequality we thus get
\bee
\left\|\eps^{-1} \textstyle \int_\R K_0(x,y;i\lambda)\, u(y)\, dy\right\|_{p}
\leq \frac{1}{2\lambda |\eps|}\,
\|e^{-\lambda|\cdot|}\|_{1} \|u\|_{p} = \|u\|_{p}.
\eee
To control the other terms in $K_\alpha(x,y; i\lambda)$, we write
\begin{align}
K^1_\alpha(x,y;i\lambda) +K^2_\alpha(x,y;i\lambda) & = \frac{-\alpha
\, e^{-\lambda |x+a|} e^{-\lambda|y+a|}}{(2\lambda+\alpha )^2 -
\alpha^2 e^{-4\lambda a}}
\label {k12} \\
&  \nonumber + \left \{
\begin {array}{ll}
\frac{\alpha^2\, e^{-\lambda |x+a|}
e^{-\lambda|y+a|}}{2\lambda((2\lambda+\alpha )^2 - \alpha^2
e^{-4\lambda a})}\, \, \left(1- e^{-2\, \lambda\, a}\,
e^{-\lambda(|y-a|-|y+a|)}\right) &  \  y<0  \nonumber \\
   &   \nonumber \\
\frac{\, \alpha^2\, e^{-\lambda |x+a|}
e^{-\lambda|y-a|}}{2\lambda((2\lambda+\alpha )^2 - \alpha^2
e^{-4\lambda a}) }\, \, \left(e^{-\lambda(|y+a|-|y-a|)}- e^{-2\,
\lambda\, a}\right)   & \ 0 \leq y
\end {array}
\right.
\end{align}
see equation  \eqref{resolvent-bis}.  It is easily seen that the function
$\lambda/((2\lambda+\alpha )^2 - \alpha^2 e^{-4\lambda a})$
is bounded for $\lambda\in(0,\infty)$. On the other hand, for $y<0$ we have $0\leq
|y-a|-|y+a| \leq 2a$. Hence for some constant $C>0$, independent of $y<0$, it holds
\bee
\sup_{\lambda>0}\, \frac{1- e^{-2\, \lambda\, a}\,
e^{-\lambda(|y-a|-|y+a|)}}{(2\lambda+\alpha )^2 -
 \alpha^2 e^{-4\lambda a}} \, \leq \, \sup_{\lambda>0}\, \frac{1- e^{-4\, \lambda\, a}}
{(2\lambda+\alpha )^2
 - \alpha^2 e^{-4\lambda a}} \, \leq C.
\eee
The last term in \eqref{k12}, which corresponds to $y\geq 0$ is
estimated in the same way. We can thus conclude that there exists a
constant $C_{\alpha,a}$, independent of $x,y$ and $\lambda$, such
that
\begin{equation} \label{estim-1}
| K^1_\alpha(x,y;i\lambda) +K^2_\alpha(x,y;i\lambda) | \, \leq \, C_{\alpha,a}\,
\frac{1}{\lambda}\, e^{-\lambda|x+a|}\, \left(e^{-\lambda|y+a|}+e^{-\lambda||y|-a|} \right).
\end{equation}
From the H\"older and Minkowski inequality we then get
\begin{align}
\|\textstyle \int_\R (K^1_\alpha(x,y;i\lambda)
+K^2_\alpha(x,y;i\lambda)) u(y)\, dy\|_{p}& \leq
\frac{C_{\alpha,a}}{\lambda}\, \, \|e^{-\lambda|\cdot|}\|_{p}\,
\left\|\textstyle\int_\R
(e^{-\lambda|y+a|}+e^{-\lambda||y|-a|}) |u(y)|\, dy\right\|_{\infty}  \nonumber \\
&  \leq \, \frac{C_{\alpha,a}}{\lambda}\, \, \|e^{-\lambda|\cdot|}
\|_{p}\, \|e^{-\lambda|y+a|}+
e^{-\lambda||y|-a|}\|_{r} \, \|u\|_{p} \nonumber \\
&  \leq \, \frac{C_{\alpha,a}}{\lambda^2}\, \, 2^{2+\frac1r}\,
p^{-\frac 1p}\, r^{-\frac 1r}\, \, \|u\|_{p}, \label{hoelder}
\end{align}
where $\frac 1p+\frac 1r =1$.  Since $K^3_\alpha(x,y;i\lambda)
+K^4_\alpha(x,y;i\lambda) = K^1_\alpha(-x,-y;i\lambda)
+K^2_\alpha(-x,-y;i\lambda)$, we arrive at
\begin{equation} \label{cont}
\frac{1}{|\eps|}\, \sum_{j=1}^4 \, \|\textstyle \int_\R
K^j_\alpha(x,y;i\lambda) u(y)\, dy \|_{p} \, \leq\, \, \mbox{const\,
}\|u\|_{p}.
\end{equation}
This completes the proof in the case $\alpha>0$.

\noindent For $\alpha\leq0$ we redefine the parameter $\lambda$ in the following way:
\bee
\lambda = -i \sqrt{\eps^{-1}+E_1(a,\alpha)}\, , \qquad \lambda > \sqrt{|E_1(a,\alpha)|}\, ,
\eee
where $E_1(a,\alpha)<0$ is the lowest eigenvalue of $H_\alpha$ defined in (\ref {Thm2a}).
Then the integral kernel of $(\eps\, T_\alpha-1)^{-1}$ is again equal to
$\eps^{-1} K_\alpha(x,y;i\lambda)$.
Moreover, there exists a constant  $C^1_{\alpha,a}$, such that
\bee
|\eps^{-1} K^1_\alpha(x,y;i\lambda) | \, \leq \, \frac{C^1_{\alpha,a}}{\lambda\, |\eps|\,
|\lambda-\sqrt{|E_1(a,\alpha)|} |} \,  e^{-\lambda|x+a|}\, e^{-\lambda|y+a|},
\eee
and similarly for $K^j_\alpha(x,y;i\lambda),\, j=2,3,4$. The
H\"older inequality applied in the same way as in \eqref{hoelder}
then gives
\bee
\frac{1}{|\eps|}\, \sum_{j=1}^4 \, \|\textstyle \int_\R
K^j_\alpha(x,y;i\lambda) u(y)\, dy \|_{p} \, \leq\,
\frac{\mbox{const\, } }{\lambda^2\, |\eps|\, |\lambda-\sqrt{|E_1(a,\alpha)|} |}  \|u\|_{p}.
\eee
Since
\bee
\lambda^2\, |\eps|\, |\lambda-\sqrt{|E_1(a,\alpha)|} | \, \geq \,
\sqrt{|E_1(a,\alpha)|}\,  \qquad \forall\,  \lambda > \sqrt{|E_1(a,\alpha)|}\, ,
\eee
the proof is completed.
\end{proof}

\begin{proof}[Proof of Theorem \ref{nls:thm}]
The Strichartz estimate \eqref{str:eq} and a straightforward application of the proof
of \cite[Prop. 4.2.1]{C} imply the uniqueness, in $H^1(\R)$, of a solution to the integral equation
\begin{equation}  \label{int-T}
\varphi_t = e^{-i t T_\alpha}\, \varphi_0 -i\nu \int_0^t e^{i(s-t)
T_\alpha}\, |\varphi_s|^2\, \varphi_s\, ds.
\end{equation}
Moreover, by Proposition \ref{aux} and \cite[Thms 3.7.1]{C} we get
the local existence of the solution to \eqref{int-T} and  the
conservation of energy and charge associated to this equation:
\begin{equation} \label{consev-T}
{\mathcal E} [\varphi_t ] + \Theta(-\alpha)\, E_1(a,\alpha)
\|\varphi_t\|^2 = {\mathcal E} [\varphi_0 ] + \Theta(-\alpha)\,
E_1(a,\alpha) \|\varphi_0\|^2 , \quad \|\varphi_t\|^2
=\|\varphi_0\|^2
\end{equation}
for $t\in (0,T)$.  This immediately implies the local existence and
uniqueness of the solution of
\begin{equation} \label{int:eq}
\psi_t = e^{-i t H_\alpha}\, \psi_0 -i\nu \int_0^t e^{i(s-t)
H_\alpha}\, |\psi_s|^2\, \psi_s\, ds
\end{equation}
by setting $\psi_t= e^{i E_1(a,\alpha) t}\, \varphi_t$ if $\alpha<
0$ and $\psi_t= \varphi_t$ otherwise. Since \eqref{int:eq} is
equivalent to \eqref{nls}, this proves the first part of the
statement. The conservation of the charge and energy, equation
\eqref{energy-consev}, is a direct consequences of \eqref{consev-T}.
\end{proof}


\appendix

\section {}
\label{technical}

\begin {lemma} \label {LemmaA} Let $n\in\N$, $\gamma>0,\, \delta \not= 0$,
and consider the following integral
\be
A_{2n+1} (\gamma , \delta ) = \int_{\R} e^{-i r^2} (r+\gamma +
i \delta )^{-2n-1} dr. \label {eqint}
\ee
\begin {itemize}
\item [i.] If $\delta >0$:
\begin{align}
A_{2n+1} & = - i (-2i)^n \sqrt {2\pi} e^{-i(\gamma + i \delta)^2/2}
D\left [ -2n-1 ,  (1-i) (\gamma + i\delta )\right ] \label
{formula10bis}
\end{align}
\item [ii.] If $\delta <0$:
\begin{align}
A_{2n+1} & = - i (-2i)^n \sqrt {2\pi} e^{-i(\gamma + i \delta)^2/2}
D\left [ -2n-1 ,  (1-i) (\gamma + i\delta )\right ]\nonumber \\
& \qquad + i^n \frac {2\pi i }{(2n)!} e^{-i (\gamma + i \delta)^2}
H_{2n} \left ( -\sqrt {i} (\gamma + i \delta) \right )\label
{formula10ter}
\end{align}
where $D(n,z)$ is the parabolic cylinder function and where $H_m$ is the
Hermite's polynom of degree $m$.

\end {itemize}

\end {lemma}

\begin {proof} Let $a=-\gamma - i \delta$, where $\delta >0$, and $z=r-a$,
then integral (\ref {eqint}) takes the form
\be
A_{2n+1} = e^{-ia^2} \int_{\R +i \delta} e^{-iz^2 - 2iza }z^{-2n-1} dz \label {eqres1}
\ee
By means of the change of variable $ s = e^{-i\pi/4} z$ and by the Cauchy theorem we then
get
\bee
A_{2n+1} = e^{-ia^2} e^{-i n\pi/2} \int_{(\ell )} e^{s^2 -2ie^{i\pi /4} a s } s^{-2n-1} ds
= e^{-ia^2} e^{-i n\pi/2} \int_{-(\epsilon )} e^{s^2 -2ie^{i\pi /4} a s } s^{-2n-1} ds,
\eee
where $(\ell )$ is the complex path $\{ \zeta \in \C \ : \ \zeta =
(x+i\delta)e^{-i\pi /4} , \ x \in \R \}$ and where $(\epsilon )$ is
the complex path defined in Figure 19.3 in \cite {AbSt}. Finally, we
set $q=\sqrt {2} s $ and apply formula (19.5.4) by \cite {AbSt} to obtain
\bee
A_{2n+1} &=& - e^{-ia^2} (-i)^n 2^n \int_{(\epsilon )} e^{q^2/2 -\left
 ( \sqrt {2} ie^{i\pi /4} a \right ) q } q^{-2n-1} dq \\
&=&  - e^{-ia^2} (-2i)^n \sqrt {2\pi } i e^{-\left ( \sqrt {2} ie^{i\pi /4} a
\right )^2/4} D \left [ -2n-1 ,\sqrt {2} ie^{i\pi /4} a \right ].
\eee
In the case $\delta <0$ then integral (\ref {eqres1}) can be written as
\bee
A_{2n+1} &=& e^{-ia^2} \int_{\R -i |\delta |} e^{-iz^2 - 2iza }z^{-2n-1} dz \\
&=& e^{-ia^2} \int_{\R +i |\delta |} e^{-iz^2 - 2iza }z^{-2n-1} dz + 2\pi i
e^{-ia^2} \mbox {Res} \left [ e^{-iz^2 - 2iza }z^{-2n-1} , z=0 \right ]
\eee
where the integral is computed as the case of $\delta
>0$ and where the residue is equal to
\bee  \mbox {Res} \left [ e^{-ir^2 }(r-a)^{-2n-1} , r=a \right ] &=&
\frac {1 }{(2n)!} \left . \frac {d^{2n}}{dr^{2n}} \left ( e^{-ir^2}
\right ) \right |_{r=a} =  \frac {i^n }{(2n)!} \left . \frac
{d^{2n}}{ds^{2n}} \left
( e^{-s^2} \right ) \right |_{s=\sqrt {i} a}  \\
&=&  \frac {i^n }{(2n)!} e^{-i a^2} H_{2n} \left ( \sqrt {i} a
\right ). \eee
\end {proof}

\begin {remark}
By formulas (19.4.6) and (19.13.1) of \cite {AbSt} equations (\ref {formula10bis})
 and (\ref {formula10ter}) can be written as
\be A_{2n+1}  = - \mbox {sign} (\delta )\, i\, (-2i)^n \sqrt {2\pi}
e^{-i(\gamma + i \delta)^2/2} D\left [ -2n-1 , \mbox {sign} (\delta
)   (1-i) (\gamma + i\delta )\right ] \label {formula10} \ee
\end {remark}

\begin {remark}
If we set
\bee
\chi = \frac {\gamma - \delta}{2} \quad \mbox { and }  \quad
\eta = \frac {\gamma + \delta}{2} \, , \quad \mbox{then}
\eee
\be A_{2n+1}  =- \mbox {sign} (\delta )\, i\, (-2i)^n  \sqrt {2\pi }
e^{-(\chi + i \eta )^2}
 D\left [ -2n-1 ,  - \mbox {sign} (\delta )2 i (\chi + i\eta )\right ] \, .
 \label {formula11}
\ee
\end {remark}

\vspace{0.2cm}

\section {Proof of lemma \ref{large-time}}
\label{long-time}

\noindent Since $|U_0(t;x,y)| = 1/2\sqrt{\pi t}$, it suffices to consider
$U_j(t;x,y),\, j=1,...,4$. Let
\bee
f_\alpha (x,y;k) = \sum_{j=1}^4 L_\alpha^j (x,y;k) = 2 \alpha k q(x,y;k) + i \alpha^2 p(x,y;k),
\eee
with
\begin{align*}
q (x,y;k) &=  -e^{ik|x+a|} e^{ik |y+a|} +e^{ik|x-a|} e^{ik |y-a|}, \quad p (x,y;k) = h(x,y;k) +
h(-x,-y;k) \\
h(x,y;k) &= e^{ik|x+a|} \left [e^{i2ka}  e^{ik |y-a|} - e^{ik |y+a|}
\right ] ,
\end{align*}
and define
\bee
r_\alpha (k) = \left ( 1 + \alpha \frac {e^{i4ka}-1}{4ik} \right ) 4 i \alpha + 4 k.
\eee
With this notation the statement of the lemma will follow, if we prove
\begin {lemma} \label {LemmaA4bis}
Let
$$
V_\alpha^t (x,y) = \int_{\R} e^{-ik^2t} \, \frac {2\, \alpha\,
q(x,y;k)}{r_\alpha (k)} dk, \qquad W_\alpha^t (x,y) = \int_{\R}
e^{-ik^2t}\,  \frac {i\, \alpha^2\,  p(x,y;k)}{k\, r_\alpha (k)} dk
.
$$
Then there exists a constant $C$ such that for $t$ large enough
\begin{equation} \label{goal}
\sup_{x,y}\,  \left | V_\alpha^t (x,y) + W_\alpha^t (x,y) \right |  \leq \, C\, t^{-1/2} .
\end{equation}
\end {lemma}

\begin {proof}
By means of the change of variable
\bee
\zeta = k \sqrt {t} - \frac{z}{2\sqrt {t}}\, , \qquad z = |x+a|+|y+a|
\eee
we get
\bee
| V_\alpha^t (x,y) | \, \leq\, \frac {C}{\sqrt {t}}\, \left |
\int_{\R} e^{-i \zeta^2 } \frac {1}{r_\alpha \left ( \zeta/\sqrt {t} + z/2t \right )} d\zeta\,  \right|.
\eee
We can estimate the above integral with the help of the Cauchy
theorem for analytic functions replacing the domain of integration
with the path
\bee
\gamma = \left \{ w \in \C \ : \  w = \zeta - \delta \frac {i\, \zeta}{\sqrt {1+\zeta^2 }}
 ,\, \delta >0, \,  \,  \zeta \in \R \right \}.
\eee
Since the function $r_\alpha(k)$ has at most two zeros, both of them
having imaginary part strictly positive, it follows that for $t$
large enough $|r_\alpha (w) |$ is bounded from below by some
positive constant in the strip $|\Im w| \leq \delta$ . A direct
calculation shows that for $t$ large enough and some
$C_\delta<\infty$ we have
\bee
 \left |
\int_{\R} e^{-i \zeta^2 } \frac {1}{r_\alpha \left ( \zeta/\sqrt {t} + z/2t \right )}
d\zeta\,  \right|\, \leq\, C\,  \int_\R e^{-\frac{2\delta\, \zeta^2}{\sqrt{1+\zeta^2}}}\,
\, d\zeta \, \leq C_\delta,
\eee
which yields the desired estimate for $V^t_\alpha$:
\begin{equation}  \label{V}
\sup_{x,y}\,  \left | V_\alpha^t (x,y) \right | \, \leq \, C\, t^{-1/2} .
\end{equation}
To prove the analogue of \eqref{V} for $W^t_\alpha$, we first observe that
\bee
h (x,y;k) = 0 \ \mbox { if } \ y \ge a.
\eee
We therefore define
\begin{align*}
h_1(x,y;k) & = h(x,y;k) = e^{ik|x+a|} e^{ik |y-a|} \, 2i \sin (2ka) \quad  \mbox { if } \ y \le - a \\
h_2(x,y;k) & = h(x,y;k)=e^{ik|x+a|} \left [e^{i2ka}  e^{ik |y-a|} - e^{ik |y+a|}  \right ]
= k e^{i k |x+a|} S(y;k) \quad  \mbox { if } \,  |y|  \leq a,
\end{align*}
and consider separately the integrals
\bee
I_j(x,y,t)=\int_{\R} e^{-i k^2 t} \frac {h_j (x,y;k)}{k r_\alpha (k)} dk , \quad  j=1,2.
\eee
For the first integral we have
\bee
I_1(x,y,t) &=& \int_{\R} e^{-i k^2 t} \frac {h_2 (x,y;k)}{k r_\alpha (k)} dk =
\int_{\R} e^{-i k^2 t} \frac {e^{ikz} 2 i \sin (2ka)}{k r_\alpha (k)} dk \\
&=& \frac {4ia}{\sqrt {t}} e^{i z^2 /4t} \int_\R e^{i \zeta^2} \frac {\sin
\left [ 2 a (\zeta  /\sqrt {t} + z /2 t ) \right ]}{2 a (\zeta  /\sqrt {t} + z /2 t )}
\frac {d\zeta }{r_\alpha (\zeta /\sqrt {t} + z/2/t )} \, ,
\eee
where $z = |x+a| + |y-a|$. Note that the for any $\delta>0$ there exists a constant $C_\delta$ such that
\bee
\left | \frac {\sin \left [ 2 a (\zeta  /\sqrt {t} + z /2 t ) \right ]}{2 a
(\zeta  /\sqrt {t} + z /2 t )} \right | \, \leq\,  C_\delta  \quad  \ |\Im \zeta | \, \leq  \delta
\eee
We can thus mimic the argument of the proof of Lemma \ref {LemmaA4bis} to arrive at the estimate
\bee
|I_1(x,y,t)| \leq C/\sqrt{t}\, ,
\eee
which holds for all $t$ large enough uniformly in $x$ and $y$.
The same arguments apply to the integral
\begin{align*}
I_2(x,y,t) &= \int_{\R} e^{-i k^2 t} \frac {h_2 (x,y;k)}{k r_\alpha (k)} dk =
\int_\R e^{-ik^2 t} e^{ik |x+a|} \frac {S (y;k)}{r_\alpha (k)} d k \\
&= \frac {1}{\sqrt{t}} e^{i|x+a|^2/4t} \int_\R e^{i \zeta^2} \frac {S (y;\zeta /\sqrt {t} +
|x+a|/2t)}{r_\alpha (\zeta /\sqrt {t} + |x+a|/2t)} d \zeta
\end{align*}
since
\bee
|S (y;w) | \le C , \ \forall y \in [-a,a], \ |\Im w | \le \delta .
\eee
Hence we can conclude that  for $t$ large enough
\begin{equation}  \label{W}
\sup_{x,y}\,  \left | W_\alpha^t (x,y) \right | \, \leq \, C\, t^{-1/2} .
\end{equation}
\end {proof}


\noindent{\bf Acknowledgements}  A.S. is grateful to R.Adami for
useful discussions.  H.K. was supported by the German Research Foundation
(DFG) under Grant KO 3636/1-1.


\end {document}